\documentclass[prd,10pt,aps,twocolumn,nofootinbib,showpacs,floatfix,amssymb,amsmath,preprintnumbers,superscriptaddress]{revtex4-1}
\usepackage{rotating,slashed,amssymb}

\newcommand{\beqn}{\begin{eqnarray}}
\newcommand{\eeqn}{\end{eqnarray}}
\newcommand{\eq}[1]{(\ref{#1})}
\newcommand{\cL}{{\cal L}}

\newcommand{\cP}{{\cal P}}
\newcommand{\cO}{{\cal O}}

\newcommand{\cA}{{\cal A}}
\newcommand{\cD}{{\cal D}}

\newcommand{\tr}{ {\rm Tr} \, }

\newcommand{\Dirac}{\slashed \cD}

\newcommand{\lr}[1]{ \left( #1 \right) }

\newcommand{\vev}[1]{ \langle \, #1 \, \rangle }

\begin{document}

 \title{Numerical evidence of the axial magnetic effect}

  \author{V. Braguta}
   \affiliation{IHEP, Protvino, Moscow region, 142284 Russia}
   \affiliation{ITEP, B. Cheremushkinskaya str. 25, Moscow, 117218 Russia}

  \author{M. N. Chernodub}\email{On leave from ITEP, Moscow, Russia.}
\affiliation{CNRS, Laboratoire de Math\'ematiques et Physique Th\'eorique,
Universit\'e Fran\c{c}ois-Rabelais Tours,\\ F\'ed\'eration Denis Poisson, Parc
de Grandmont, 37200 Tours, France}
\affiliation{Department of Physics and Astronomy, University of Gent, Krijgslaan
281, S9, B-9000 Gent, Belgium}

  \author{K. Landsteiner}
     \affiliation{Instituto de F\'{\i}sica Te\'orica UAM/CSIC, C/ Nicol\'as
Cabrera 13-15,\\
     Universidad Aut\'onoma de Madrid, Cantoblanco, 28049 Madrid, Spain}

  \author{M. I. Polikarpov}\email{Deceased.}
     \affiliation{ITEP, B. Cheremushkinskaya str. 25, Moscow, 117218 Russia}
     \affiliation{Moscow Inst Phys \& Technol, Institutskii per. 9, Dolgoprudny, Moscow Region, 141700
Russia}

\author{M.V. Ulybyshev}
\affiliation{ITEP, B. Cheremushkinskaya str. 25, Moscow, 117218 Russia}
\affiliation{Institute for Theoretical Problems of Microphysics, Moscow State
University, Moscow, 119899 Russia}

\date{July 18, 2013}
 
\begin{abstract}
The axial magnetic field, which couples to left- and right-handed fermions with
opposite signs, may generate an equilibrium dissipationless energy flow of
fermions in the direction of the field even in the presence of interactions. We
report on numerical observation of this Axial Magnetic Effect in quenched
$SU(2)$ lattice gauge theory. We find that in the deconfinement (plasma) phase
the energy flow grows linearly with the increase of the strength of the axial
magnetic field. In the confinement (hadron) phase the Axial Magnetic Effect is
absent. Our study indirectly confirms the existence of the Chiral Vortical
Effect since both these effects have the same physical origin related to the
presence of the gravitational anomaly. 
\end{abstract}
 \pacs{11.15.-q, 12.38.Mh, 47.75.+f, 11.15.Ha}
 \preprint{IFT-UAM/CSIC-13-033}
 \maketitle

Anomalies belong to the most characteristic and fundamental properties of
relativistic quantum field theories.
They signal an incompatibility between quantization and the
symmetries present at the classical level. 
While the effects of anomalies in vacuum are well understood it has only
recently been fully appreciated
that anomalies play also an extraordinary important role at finite temperature
and density.
In particular they give rise to new non-dissipative transport phenomena. The
most well-known of these
is the so-called Chiral Magnetic Effect (CME) \cite{ref:CME}, 
describing the generation of an electric current 
parallel to a magnetic field in the presence of an imbalance between the number
of right-handed and
left-handed fermions (a nice review can be found in Ref.~\cite{ref:Valya}). The CME is thought to be responsible for charge
asymmetries observed in heavy ion
collisions at RHIC and LHC~\cite{ref:CME:experiment}.
It also might play a role in the transport
properties of advanced new materials,
the so-called Weyl semi-metals in which the effective charge carriers can be
modeled as $3+1$ dimensional
Dirac fermions~\cite{ref:Weyl}.

The CME is however only one representative of a whole class of anomaly related
transport phenomena. A full classification
of such phenomena has been obtained via Kubo formulas
in~\cite{Landsteiner:2011cp}. 
It turned out
that not only the usual axial or chiral
anomalies give rise to dissipationless transport but that there is also a
distinguished place for the axial gravitational anomaly.

In general, anomaly related transport is sourced by either external magnetic
fields or by vortices in the fluid of chiral fermions \cite{Erdmenger:2008rm}\footnote{An
early  manifestation of this effect in rotating ensembles of neutrinos was found in Ref.~\cite{Vilenkin:1979ui}.}. 
Thus we can distinguish between chiral magnetic and chiral vortical effects. The
gravitational anomaly comes in through the chiral
vortical effect (CVE). Even in the absence of chemical potentials the
gravitational anomaly gives rise to a chiral vortical effect
at finite temperature
\begin{equation}\label{eq:cve}
 \vec{J}_5 = \sigma \vec\omega\,,
\end{equation}
where $\vec{J}_5$ is the axial current and ${\vec \omega} = \nabla\times \vec{v}$ is the vorticity of the fluid velocity
$\vec{v}$. In the absence of matter, the conductivity
\beqn\label{eq:sigma}
\sigma = \left( \sum_l q_l - \sum_r q_r  \right) \frac{T^2}{24}\,,
\eeqn
depends on the temperature $T$ and the gravitational anomaly coefficient. 
Equation~\eq{eq:cve} is valid for a theory consisting of massless
fermions. In a basis of left- and right-handed Weyl fermions $q_l$ ($q_r$) are
the charges of the left-handed (right-handed) fermions.
This effect has been confirmed at 
strong coupling via the gauge-gravity correspondence in 
Ref.~\cite{Landsteiner:2011iq}. 

The chiral vortical effect~(\ref{eq:cve}) is quite remarkable since,
contrary to the CME, it does not rely on the explicit 
introduction of an axial chemical potential: in Eq.~(\ref{eq:cve}) 
the difference of the right- and left-handed charges in Eq.~\eq{eq:sigma} corresponds to the vacuum content of the theory.
On the other hand, its relation to the axial gravitational anomaly is at the moment still 
somewhat less direct than the relation of the CME to the axial anomaly. Purely
hydrodynamic arguments such as in 
Ref.~\cite{Son:2009tf} are not able to fix the 
prefactor in the conductivity~\eq{eq:sigma} 
due to a mismatch in the order of derivatives in which the gravitational anomaly can
apparently influence hydrodynamics (see however \cite{Jensen:2012kj}). 

It has further been proven that the transport law (\ref{eq:cve})
does not get renormalized in perturbation theory in theories 
which contain fermions and scalars only.
If there are however dynamical gauge fields that contribute to the axial anomaly
then
it was shown that a non-vanishing two loop contribution arises
\cite{Golkar:2012kb}. 
In QCD this is of course the case. The axial anomaly has a gluonic contribution
and therefore
one expects a strong renormalization of the chiral vortical effect in QCD.   

Since the CVE is a dissipationless and stationary effect it is
accessible via Euclidean field theory
and can, in principle, be studied numerically in simulations of lattice gauge
theories. A straightforward implementation of rotating fluid
on a lattice seems, however, a rather non-trivial task 
due to obvious incompatibility of the small, discrete rotational symmetry group
of the Euclidean lattice with smooth, continuous rotations used in
Eq.~\eq{eq:cve}.
Fortunately there is an alternative way of accessing this particular transport
phenomenon that is well suited for implementation on the lattice. 

In order to compute the CVE for the axial current one could make use of the Kubo
formula
\begin{equation}\label{eq:kubo}
 \sigma = - \lim_{p_j \to  0} \frac{i}{2p_j}\sum_{i,k} \epsilon_{ijk}\left\langle J^i_5 T^{0k}\right\rangle\,,
\end{equation}
where $J^i_5$ are spatial components of the axial current and $T^{0k}$ are
temporal-spatial
components of the energy-momentum tensor. Since the correlator is to be
evaluated at
zero frequency one can reverse the order of the operators 
$J^i_5$ and $T^{0k}$ in Eq.~\eq{eq:kubo}
and obtain a
new
effect corresponding to the generation of an energy current $J_\epsilon^k =
T^{0k}$ in
the background field that couples to the axial current, this is an axial
magnetic field.
One finds therefore that the chiral vortical conductivity~\eq{eq:sigma} also
appears in the
new transport formula~\cite{Landsteiner:2011cp}:
\begin{equation}\label{eq:ame}
 \vec{J}_\epsilon = \sigma \vec{B}_5\,,
\end{equation}
which represents the Axial Magnetic 
Effect (AME). The transport law~\eq{eq:ame} describes the generation of an
equilibrium dissipationless energy current in the presence of an axial magnetic field at
finite temperature. 
Note that a priori equation (\ref{eq:ame}) is valid only for weak axial magnetic field since
it was derived via linear response theory. Our numerical results show that the linear
behavior in $B_5$ is valid even away from the weak field limit.

The practical advantage of the AME formula~\eq{eq:ame} is that the axial
magnetic field ${\vec B}_5$ can be relatively easy implemented on the Euclidean
lattice, while the implementation of the vorticity $\vec \omega$ is a much more
difficult task. On the other hand, both the CVE and AME have the same physical
nature -- which is also clear from the very fact that they share the same
conductivity coefficient~\eq{eq:sigma}
-- originating due to the presence of the gravitational
anomaly~\cite{Landsteiner:2011cp}. 
Thus, in this article, we concentrate on numerical evaluation of the AME
law~\eq{eq:ame} in the context of the quenched SU(2) lattice gauge theory for
three different temperatures, which represent three basic regions of the phase
diagram: the deconfinement regime, the critical confinement-deconfinement
region, and the confinement phase. 

The important feature of the AME is that it is realized in the pure vacuum with
all chemical potentials equal to zero, $\mu = \mu_5 = 0$. The dissipationless
equilibrium energy flow is achieved at finite temperature, $T \neq 0$, in the
presence of the axial magnetic field ${\vec B}_5$, which distinguishes
left-handed and right-handed quarks. The axial magnetic field couples to the
left-handed and right-handed quarks with opposite charges respectively.

In order to check the existence of the AME law~\eq{eq:ame} it is sufficient to
consider one type of fermion with a unit charge:
\beqn
q^L_{5} = - q^R_{5} = + e\,.
\label{eq:RL:charges}
\eeqn
The coupling of the quarks to the chiral field $A^5_\mu$ is described by the
following Lagrangian:
\beqn
\cL_5 = {\bar \psi} (\partial_\mu - i g A^a_\mu t^a - i \gamma^5 e A_{5,\mu})
\gamma^\mu  \psi \equiv {\bar \psi} \Dirac_5 (A_5)\psi\,, \quad
\label{eq:D5f}
\eeqn
where $A^a_\mu$ is the nonabelian $SU(N)$ gauge field and $t^a$ are the
generators of the $SU(N)$ gauge group.

The energy flow in the transport law~\eq{eq:ame} is given by the expectation
value of the off-diagonal component of the stress-energy tensor $T^{\mu\nu}$,
\beqn
 J^i_\epsilon & = & \vev{T^{0i}} \equiv \frac{i}{2} \vev{{\bar \psi}  (\gamma^0
\cD^{i}_5 + \gamma^i \cD^{0}_5)\psi }\,,
\label{eq:J:E}
\eeqn
where the latin index $i = 1,2,3$ labels the spatial coordinates and $\mu=0$ is
the time direction.

We introduce the stationary uniform axial magnetic field in the third direction
$B_{5,i} = B_5 \cdot \delta_{i,3}$ by setting $A_{5,0} = A_{5,0} = 0$, $A_{5,1}
= - x_2 B_5/2$ and $A_{5,2}  = x_1 B_5/2$.

The energy flow can be implemented on the lattice via a straightforward discretization 
of Eq.~\eq{eq:J:E} using the linear combinations of the nonlocal lattice correlator,
\beqn
C_\mu (x,y;A) & = & \vev{\bar{\psi}  (x) U_{x,y}(A^a_\mu) \gamma_{\mu} {\psi} (y)}_A \nonumber \\
& \equiv & \tr\lr{U_{x,y}(A^a_\mu) \, \frac{1}{\mathcal{D}_{5} + m} \, \gamma_{\mu}}_{x,y;A}\,.
\label{eq:core:latt}
\eeqn
In this formula, the expectation value is taken over the fermion field in a fixed background of non-Abelian $A^a_\mu$ and axial $A_{5,\mu}$ fields,
and the trace is taken over color and spinor indices.
Here $U_{x,y}$ is the gluon string between the lattice points $x$ and $y$ which makes Eq.~\eq{eq:core:latt} gauge invariant.
The Dirac operator $\cD_5$ is given in Eq.~\eq{eq:D5f}. 

We calculate the correlation functions~\eq{eq:core:latt} numerically, using
lattice Monte-Carlo simulations of quenched $SU(2)$ lattice gauge theory
following numerical setup
of Refs.~\cite{Buividovich:2009wi}. The quark fields are
introduced by the overlap lattice Dirac operator $\mathcal{D}$ with exact chiral
symmetry~\cite{Neuberger:98:1}. 
The correlation functions~\eq{eq:core:latt} are substituted in the discretized
version of Eq.~\eq{eq:J:E} and then the whole expression is averaged over an
equilibrium ensemble of finite temperature configurations of non-Abelian gauge
fields $\cA_\mu$,
\begin{eqnarray}
\left\langle \cO \right\rangle = \left(\int D A^a_{\mu}\,e^{-S_{YM} [A^a_{\mu}]
}\right)^{\hskip -0.5mm -1} \hskip -1.5mm  \int D A^a_{\mu}\,e^{-S_{\mathrm{YM}}
[A^a_{\mu}] }\, \cO\,,
\quad
\nonumber
\end{eqnarray}
where $S_{\mathrm{YM}}(A^a_{\mu})$ is the lattice action for the gluons
$A^a_{\mu}$. 

There are at least two ways to calculate the fermion
propagator~\eq{eq:core:latt} in the presence of the axial magnetic field in a
finite-volume lattice. One can introduce the axial magnetic field
straightforwardly by modifying the spatial boundary conditions for fermions
according to the general approach of Ref.~\cite{Wiese:08:1}. Alternatively, one
can make use of the identity,
\beqn
& & {\mathrm{tr}} \left[S_5(A_5) \, \gamma_{\mu}\right] \equiv {\mathrm{tr}} \left[ (\cP_R + \cP_L) S_5(A_5) \, \gamma_{\mu}\right] \nonumber \\
& & \quad\, = {\mathrm{tr}} \left[\cP_R \, S(A_5) \gamma_{\mu}\right] + {\mathrm{tr}} \left[\cP_L \, S(-A_5) \gamma_{\mu}\right], \qquad
\label{eq:identity}
\eeqn
where $\cP_{R,L} = (1 \pm \gamma^5)/2$ are the right and left chiral projectors, the trace is taken over spinor indices and 
\beqn
S_5(A_5) =  \left[\Dirac_5 (A_5)\right]^{-1}, \qquad S(\cA) & = & \left[\Dirac (\cA)\right]^{-1},
\eeqn
are the Dirac operators for the massless fermions in the background of the axial
field $A_{5,\mu}$ and the usual Abelian gauge field $\cA$, respectively. The
former is defined in Eq.~\eq{eq:D5f} while the latter has the usual form:
\beqn
\cD_\mu(\cA) = \partial_\mu - i g A^a_\mu t^a - i e \cA_\mu\,.
\eeqn
It is worth stressing that in the right hand side of Eq.~\eq{eq:identity} the axial
gauge field $A_5$ appears as the Abelian field with opposite signs for
right-handed and left-handed fermions. This is an expected property given the
very definition of the axial magnetic field.

Identity~\eq{eq:identity} is valid regardless of the dynamical
generation of quark mass and chiral symmetry breaking 
since this identity is a generic property of the fermion
operator itself while the mentioned phenomena are the particular properties of
the expectation values of this operator.

Thus, identity~\eq{eq:identity} allows us to express the energy
flow~\eq{eq:J:E} of the massless fermions via the standard Dirac operator in a
background of usual magnetic field. This property is particularly useful for
numerical calculations with overlap fermions since we can use the already
existing techniques of Ref.~\cite{Buividovich:2009wi}.

We evaluate the energy flow~\eq{eq:J:E} in the quenched $SU(2)$ gauge theory
using 300 configurations of the gluon gauge field for each value of the
background axial magnetic field. 
Theoretically, the presence of the vacuum fermion loops is neither crucial nor necessary for the anomalous transport phenomena~\cite{Landsteiner:2011cp}
so that we expect that the quenching should give a minor effect. We also expect that the reduced number of colors (2 colors instead of 3) may affect the numerical value on the slope $\sigma$ of the anticipated linear behavior~\eq{eq:ame}.

We consider the asymmetric lattices $L_s^3 L_t$
with three temporal lengths $L_t = 4,6,8$ and the fixed spatial length $L_s =
14$. We use the improved lattice action for the gluon fields with the lattice
coupling $\beta=3.2810$ which corresponds to the lattice spacing~$a =
0.103\,\mathrm{fm}$~\cite{Bornyakov:2005iy}.

Similarly to the usual magnetic field, the axial magnetic field is quantized due
to the periodicity of the gauge fields in a finite volume :
\beqn
B_5 = k \, B_{5,\mathrm{min}}\,, \qquad 
e B_{5,\mathrm{min}} = \frac{2 \pi}{L^2_s} \approx 0.117\, {\mathrm{GeV}}^2\,,
\qquad
\label{eq:B:quant}
\eeqn
where the integer $k = 0, 1, \dots, L_s^2/2$ determines the number of elementary
magnetic fluxes which pass through the boundary of the lattice in the
$(x^1,x^2)$ plane. 
Notice that the elementary (minimal) field~\eq{eq:B:quant} in our simulation is
three times weaker compared to the field used in our previous studies,
Ref.~\cite{Buividovich:2009wi}, because in the present paper the fermion field
carries a unit electric charge.

The maximal possible value of the quantized flux, $k = L^2_s/2 = 98$,
corresponds to an extremely large magnetic field. In this case the magnetic
length $L_B \sim (e B)^{-1/2}$ is the order of the lattice spacing, $L_B \sim
a$. In order to avoid possible ultraviolet artifacts, we consider relatively
weak axial magnetic fields with $k = 0, 1, \dots k_{\mathrm{max}}$, where the
flux number is limited by $k_{\mathrm{max}} = 10 \ll l^2/2$, so that our
strongest magnetic field is $e B_{\mathrm{max}} \approx 1.17\,
{\mathrm{GeV}}^2$.

In order to increase the efficiency of our numerical algorithm we introduce a
small bare quark mass $m \sim 20\,\mathrm{MeV}$ in the overlap fermion operator~\eq{eq:core:latt}.
Despite this bare mass being very small, we have carefully checked the
applicability of Eq.~\eq{eq:identity} to our numerical setup by making sure that
the energy flow is insensitive to the variations of the quark mass. For example,
a two-fold increase of the mass leads to a less than $1\%$ change of the central
values of our observable (this is much smaller than our statistical errors).

In $SU(2)$ gauge theory the critical deconfinement temperature is $T_c =
303\,\mathrm{MeV}$. Our three lattices with temporal extensions $L_t = 4,6,8$
correspond, respectively, to the deconfinement phase ($T = 1.58\, T_c$), the
vicinity of the confinement--deconfinement phase transition ($T = 1.05\, T_c$)
and the confinement phase ($T = 0.79\, T_c$).
\begin{figure}[!thb]
\begin{center}
\includegraphics[scale=0.55,clip=false]{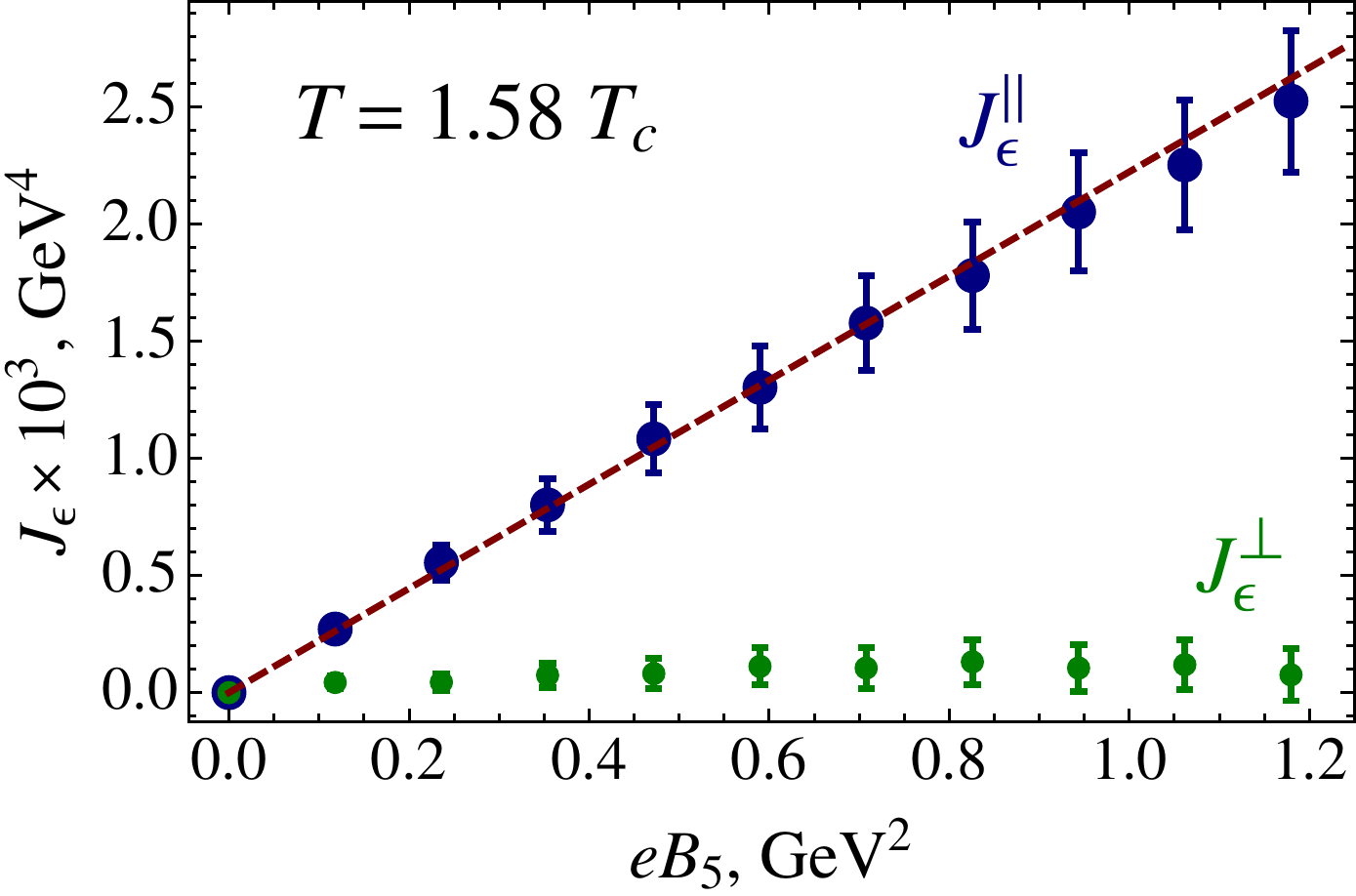} 
\end{center}
\caption{Energy flow~\eq{eq:J:E} parallel ($J^\|_\epsilon$) and perpendicular
($J^\perp_\epsilon$) to the direction of the axial magnetic field $B_5$ in the
deconfinement phase at $T = 1.58 \, T_c \equiv 479\,\mathrm{MeV}$. The red
dashed line represents the best linear fit confirming the existence of the Axial
Magnetic Effect~\eq{eq:ame}. The slope of the fit is given by the
conductivity~\eq{ref:conductivity}.}
\label{fig:result}
\end{figure}

In Fig.~\ref{fig:result} we show the energy flow~\eq{eq:J:E} as a function of
the axial magnetic field $B_5$ in the deconfinement phase at $T = 1.58 \, T_c$.
One can see that the energy flow parallel to the axial magnetic field $J^\|_\epsilon$ is a
linear function of the field strength as predicted by the AME transport
law~\eq{eq:ame}. The energy flow in the transverse directions $J^\perp_\epsilon$ is zero within the
error bars, as expected. 

\begin{figure}[!thb]
\begin{center}
\includegraphics[scale=0.6,clip=false]{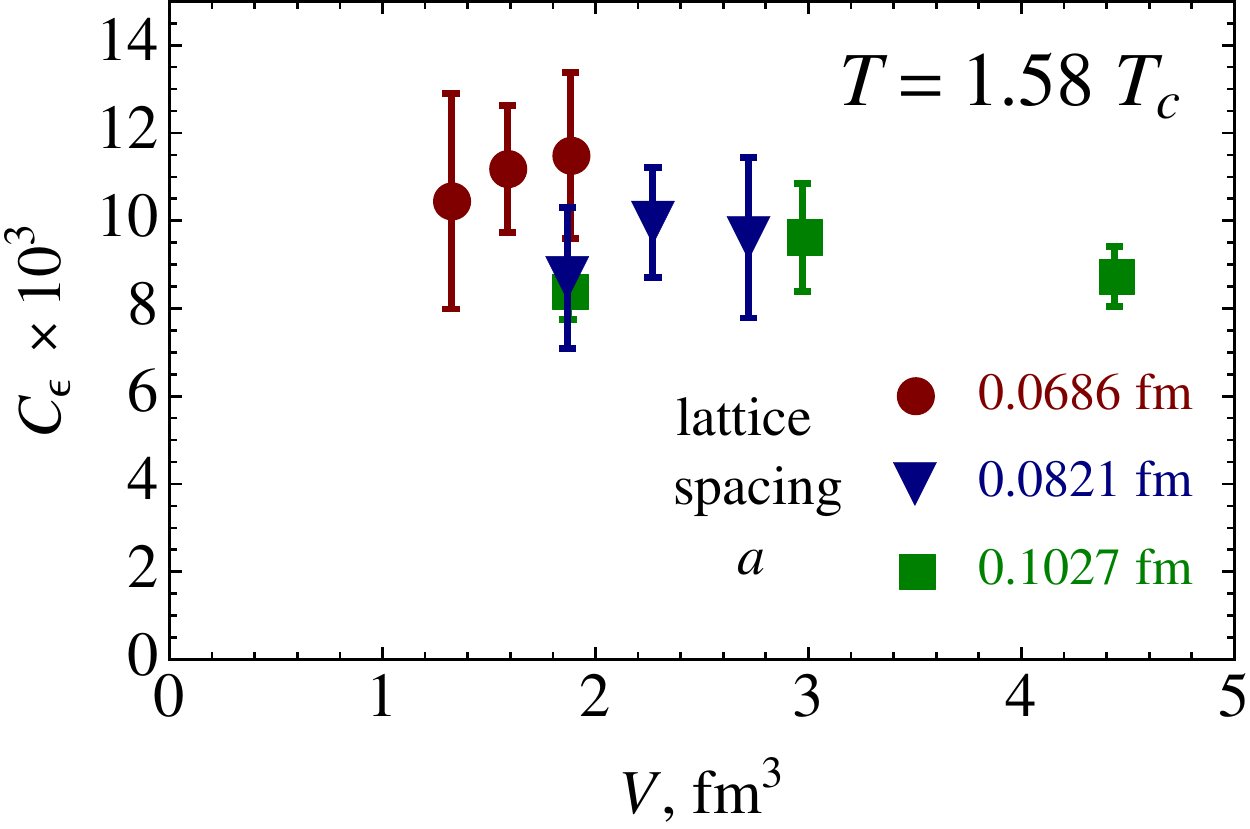} 
\end{center}
\caption{The dimensionless prefactor $C_{\epsilon} = J_{\epsilon}/(eB_5 T^2)$ as the function of the volume $V$ at three different values of the lattice spacing $a$. The expected theoretical value is $C_\epsilon = 1/6$.}
\label{fig:volume}
\end{figure}

In Fig.~\ref{fig:volume} we demonstrate that our result does not depend (within reasonable error bars) on the volume of the system being simultaneously robust with respect to the variations of the lattice spacing (the latter plays a role of the inverse ultraviolet cutoff). To this end we have performed simulations on the lattices in the range $L_s = 12 \dots 18$ and $L_t = 4 \dots 8$ at $\beta = 3.281,\ 3.398$ and $2.5$. We have also checked numerically that the usual magnetic field does not induce any energy flow.

The linear fit of the energy flow -- shown by the dashed line in Fig.~\ref{fig:result} -- gives us the following value of the energy flow conductivity:
\beqn
\sigma(T){\biggl|}_{T = 1.58 \,T_c} = 2.22(3)\times 10^{-3} \, \mbox{GeV}^2\,.
\label{ref:conductivity}
\eeqn
For a conformal theory with two colors of fermions,  the conductivity coefficient is expected to be $\sigma_{\mathrm{theor}} = T^2/6$  which gives us one order of magnitude larger
value $\sigma_{\mathrm{theor}} \approx 0.0382\,\mathrm{GeV}^2$ at
this temperature. We expect that this large difference is not related to the absence of the fermion vacuum loops in our simulations because the latter do not play an essential role in the AME~\cite{Landsteiner:2011cp}.

We see no signature of energy flow neither in the vicinity of the phase
transition nor in the confinement phase. 

Summarizing, we have numerically observed the existence of the Axial Magnetic
Effect (AME)
in the quenched $SU(2)$ gauge theory on the lattice. We have found that the
equilibrium energy flow of massless fermions is parallel to the direction of
(and proportional to the strength of) the time-independent uniform axial
magnetic field. Our study also indirectly confirms the existence of the Chiral
Vortical Effect (CVE) since both these effects have the same physical nature
originating from the gravitational anomaly.

\acknowledgments

We thank P.~V.~Buividovich for useful discussions.
The work of the Moscow group was supported by Grant "Leading Scientific
Schools" No. NSh-6260.2010.2, RFBR-11-02-01227-a, RFBR-12-02-31249,
RFBR-13-02-01387, Federal Special-Purpose Program "Cadres" of the
Russian Ministry of Science and Education, and by a grant from the
FAIR-Russia Research Center. Numerical calculations were performed at
the ITEP computer systems ``Graphyn'' and ``Stakan''. The work of M.N.C.
was partially supported by grant No. ANR-10-JCJC-0408 HYPERMAG of
Agence Nationale de la Recherche (France). The work of K.L. was
partially supported by by Plan Nacional de Altas Energ\'\i as FPA
2009-07890, Consolider Ingenio 2010 CPAN CSD200-00042 and
HEP-HACOS S2009/ESP-2473. We thank the ECT* center in Trento, 
Italy, where this collaboration has started during the
``Workshop on QCD in strong magnetic fields'' in November 2012. M.N.C.
is grateful to Instituto de F\'{\i}sica Te\'orica UAM/CSIC in Madrid
for kind hospitality during his visit.

\end{document}